\documentclass[a4paper,12pt]{article}

\usepackage[english]{babel}
\usepackage[T1]{fontenc}

\usepackage[onehalfspacing]{setspace}

\usepackage[a4paper,top=3cm,bottom=3cm,left=2.5cm,right=2.5cm]{geometry}

\usepackage{amsmath}
\usepackage{graphicx}
\usepackage[colorlinks=true, allcolors=blue]{hyperref}
\usepackage{textcomp}
\usepackage{authblk}
\usepackage{lmodern}

\usepackage{cite}

\title{Dynamic Growth/Etching Model for the Synthesis of Two-Dimensional Transition Metal Dichalcogenides via Chemical Vapour Deposition}
\author[1,*]{Erik Pollmann}
\author[1]{André Maas}
\author[1]{Dave Marnold}
\author[1]{Alfred Hucht}
\author[1,2]{Rahel-Manuela Neubieser}
\author[1,3]{Mike Stief}
\author[1]{Lukas Madau\ss}
\author[1]{Marika Schleberger}
\affil[1]{Faculty of Physics and CENIDE, University of Duisburg-Essen, Lotharstra\ss e 1, D-47057 Duisburg, Germany}
\affil[2]{current affiliation: Fraunhofer Institute for Microelectronic Circuits and Systems, Finkenstra\ss e 61, D-47057 Duisburg, Germany}
\affil[3]{current affiliation: NanoFocus AG, Max-Planck-Ring 48, D-46049 Oberhausen, Germany}

\affil[*]{corresponding authors: erik.pollmann@uni-due.de, marika.schleberger@uni-due.de}

\begin{document}
\maketitle

\begin{abstract}

The preparation of two-dimensional transition metal dichalcogenides on an industrially relevant scale will rely heavily on bottom-up methods such as chemical vapour deposition. In order to obtain sufficiently large quantities of high-quality material, a knowledge-based optimization strategy for the synthesis process must be developed. A major problem that has not yet been considered is the degradation of materials by etching during synthesis due to the high growth temperatures. To address this problem, we introduce a mathematical model that accounts for both growth and, for the first time, etching to describe the synthesis of two-dimensional transition metal dichalcogenides. We consider several experimental observations that lead to a differential equation based on several terms corresponding to different supply mechanisms, describing the time-dependent change in flake size. By solving this equation and fitting two independently obtained experimental data sets, we find that the flake area is the leading term in our model. We show that the differential equation can be solved analytically when only this term is considered, and that this solution provides a general description of complex growth and shrinkage phenomena. Physically, the dominance suggests that the supply of material via the flake itself contributes most to its net growth. This finding also implies a predominant interplay between insertion and release of atoms and their motion in the form of a highly dynamic process within the flake. In contrast to previous assumptions, we show that the flake edges do not play an important role in the actual size change of the two-dimensional transition metal dichalcogenide flakes during chemical vapour deposition.

\end{abstract}


\section{Introduction}

Scaling up the lateral extension of two-dimensional transition metal dichalcogenides (2D TMDCs) is crucial to exploit the full potential in promising applications in fields such as (opto)electronics, sensing and catalysis \cite{Radisavljevic.2011, Yin.2012, LopezSanchez.2013, Li.2012, Urban.2019, Lauritsen.2003, Jaramillo.2007, Le.2014, Li.2016b, Ye.2016, Yin.2016, Li.2016, Dong.2018,  Madau.2018, Sun.2019}. In addition to the goal of a wafer-scale coverage of 2D TMDCs, a more advanced and challenging requirement is that the 2D TMDC is as single crystalline as possible, i.e. that it exhibits maximum large domains -- called flakes in non-closed 2D TMDC films.

By now, for the fabrication of 2D TMDCs in large scales, an access via top-down methods exists \cite{Magda.2015, Velicky.2018, Pollmann.2021}. However, in particular bottom-up methods such as chemical vapour deposition (CVD) are promising due to their potential compatibility with processes for thin film fabrication established in the semiconductor industry. With bottom-up methods, the desired materials are formed by self-assembly of the corresponding precursor atoms. Requirements for this are, for instance, suitable high temperatures and the supply of sufficient precursor material.

After Lee et al. first reported the successful growth of the 2D TMDC molybdenum disulphide (MoS$_{2}$) with CVD in 2012 \cite{Lee.2012}, various 2D TMDCs have been synthesized on different substrates, as well as on other van der Waals and 2D materials \cite{Lee.2013, Yu.2013, Zhang.2013, Huang.2014, Ling.2014, Wang.2014, Okada.2014, Wang.2014b, Ago.2015, Dumcenco.2015, Chen.2015b, Liu.2016, Chen.2015, Cain.2016, Zhu.2017, Zhou.2018, Pollmann.2018, Zhang.2019, Pollmann.2020, Okada.2020, Pollmann.2020b}. One of the most rudimentary realization of CVD for 2D TMDCs is based on the use of two solid precursor sources containing either the chalcogen (e.g. elemental sulphur powder) or the transition metal (e.g. transition metal oxides or chlorides). The similarities of the process design for many 2D TMDC species and of their resulting morphology imply identical atomic kinetics during CVD. 

In the last decade, deeper insights into the growth mechanisms of 2D TMDCs (mostly 2D MoS$_{2}$) have been collected by refining the process systems and recipes as well as by developing models. For example, the initial nucleation has been studied extensively \cite{Ling.2014, Chen.2015b, Liu.2016, Cain.2016, Zhu.2017, Pollmann.2018, Zhang.2019, Pollmann.2020}. As a result, concepts for controlling nucleation were presented, e.g. by the use of seeding promoters \cite{Ling.2014} or artificial defects in the substrate \cite{Pollmann.2018, Zhang.2019}. Furthermore, different growth rates for the 2D TMDC crystal facets have been identified as the reason for the typically (equilateral) triangular flake shape from CVD. These growth rates differ in their dependencies of the ratio of the precursor atom species (transition metal or chalcogen), allowing potentially to control of the edge termination and even the shape of the resulting flakes by the precursor atom concentrations in the gas phase \cite{Wang.2014b, Zhu.2017}. Recently, concepts have been proposed in order to describe the dynamics and the stability of the orientation of 2D TMDC flakes growing on crystalline van der Waals materials \cite{Okada.2020}. Here, the flake orientation might be controlled by substrate defect engineering \cite{Zhang.2019b, Zhang.2019}.

However, one phenomenon has yet only been insufficiently elucidated. It is experimentally found that 2D TMDC flakes first grow and then shrink again as the process duration increases, see the publication by Chen et al. \cite{Chen.2015} or Fig.~\ref{Fig:ExperimentalData}. Because degradation of 2D TMDCs is also facilitated by increased temperatures\cite{Zhou.2013, Wu.2013, Yamamoto.2013, Ye.2016, Lv.2017, Maguire.2019, Yao.2020, Kaupmees.2020, Cullen.2021, Lin.2015, Chen.2018} [and see SI~1 (Fig.~S1)], a dynamic process between growth and etching during a CVD process must obviously exists. Chen et al. assume -- without providing any theoretical model -- that insertion and release of atoms only takes place at the edges of the grown 2D TMDC flake (labelled and discussed below as growth rate $G_\mathrm{1D}$ and etching rate $E_\mathrm{1D}$, respectively). This assumption seems to be straightforward and intuitive for describing synthesis of 2D TMDCs as the basal plane is chemically rather inert, while edges represent the active sites, at least at low temperatures. In addition, Chen et al. explain their observations by the absence of adsorbed material on the basal plane of growing flakes. That material supply takes place exclusively via the substrate is also implied by Wang et al. \cite{Wang.2014b}. But are these assumptions really adequate to describe the growth process?

In this paper, we introduce an advanced concept explaining the experimentally observed fact, that the size of 2D TMDC flakes at first increases and then decreases again during CVD. Our mathematical concept is based on considerations taking material supply and transport into account as well as its change over time and is complemented by thermal degradation/etching mechanisms concluded from experiments. Finally, we apply the resulting equations to data for MoS$_{2}$ on sapphire by Chen et al. \cite{Chen.2015} and to our own data for tungsten disulphide (WS$_{2}$) on sapphire. Contrary to intuitive assumption we find, both mathematically and by the best fits on both data sets, that the change in area of the 2D TMDC flakes is largely proportional to the flake area itself.

\section{Results and Discussion}

To describe the experimental results correctly, our model must account for growth and shrinking mechanisms which will depend on material transport and supply. Within the model, the time-dependent change of the area of a two-dimensional TMDC flake $\frac{\mathrm{d}A}{\mathrm{d}t}$ depends on its own, current size (defined by its area $A$ or its edge length $L$, respectively). In this context, we distinguish between rates that result in an increase in the flake size and those that lead to a decrease. Accordingly, they are called growth rate $G$ and etching rate $E$, respectively. At first, we will discuss these rates in order to develop our basic differential equation shown later in Eq.~(\ref{Eq:TotalDE}).

\begin{figure}[t]
\centering
\includegraphics[width=1.0\textwidth]{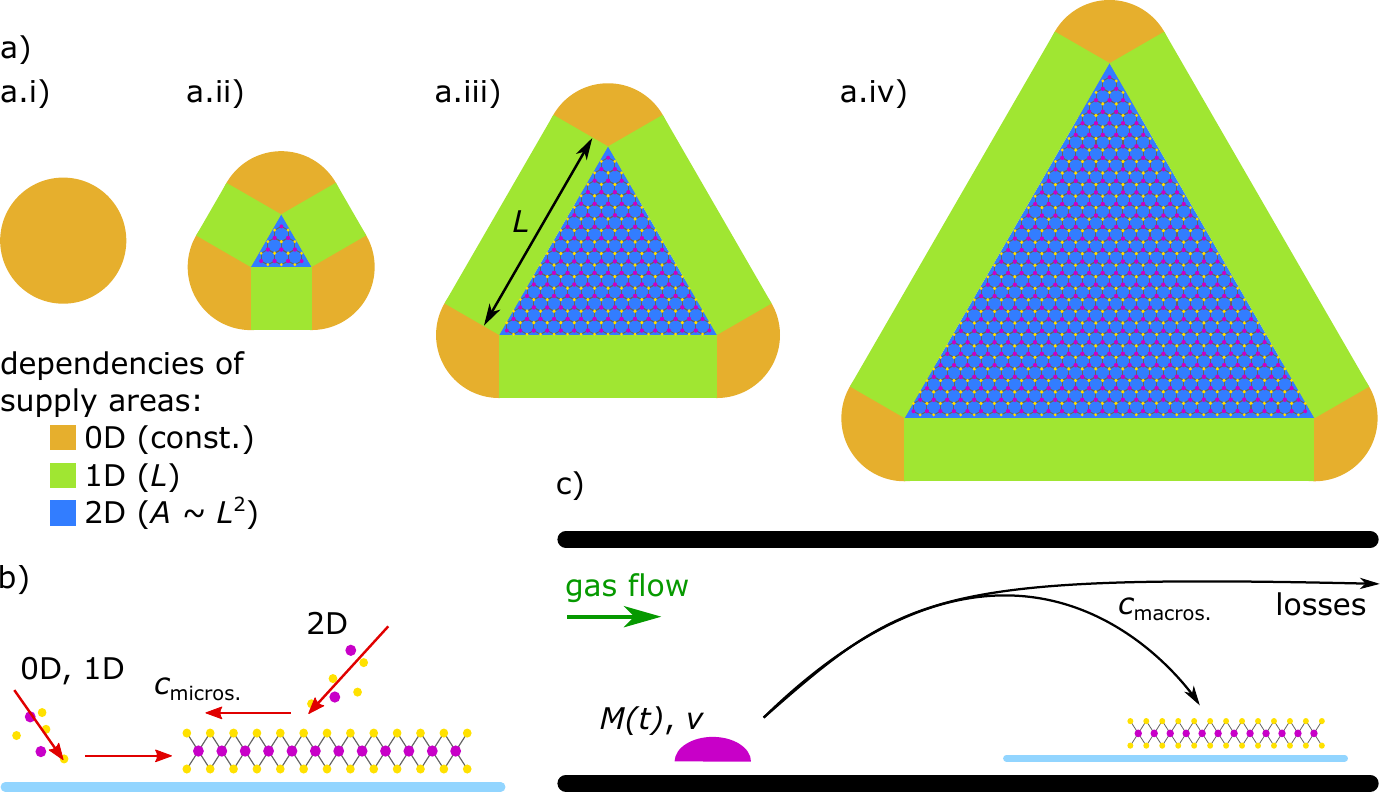}
\caption{Illustration of the growth related factors for the dynamic growth/etching model described by Eq.~(\ref{Eq:TotalDE}). a) Evolution of three different supply areas with the flake size. The size of the supply areas are either independent of spatial dimensions (orange, 0D), or dependent on one (green, 1D) or two dimensions (blue, 2D), respectively. b) Precursor atoms either diffuse on the surface of the substrate (0D, 1D) or of the flake (2D) to be inserted at the flake edge. The constant $c_\mathrm{micros.}$ summarises microscopic constants such as the adsorption and desorption rate, the diffusion constant as well as the reaction probability and differs for the supply areas. c) The schematic of a basic CVD system illustrates the dependency of the adsorbed material on the amount of source material $M(t)$ and its specific vaporisation rate $v$. The constant $c_\mathrm{macros.}$ takes into account losses caused due to the transport from the source to the target substrate.}
\label{Fig:SupplyAreas}
\end{figure}

The growth rate of a single 2D TMDC flake $G$ depends on the size of the total supply area from which precursor material can agglomerate to form a new flake (nucleation) or diffuse to appropriate sites in pre-existing flakes in order to increase its size (direct growth) or to compensate etching (indirect growth). The supply area itself is related to the flake size -- or simplified: larger flakes can ``catch'' more precursor material. Because, corresponding to Fig.~\ref{Fig:SupplyAreas}~a), the total supply area is composed of different areas with different dependencies on the flake size (and on the different surfaces, see Fig.~\ref{Fig:SupplyAreas}~b)), we distinguish between three supply areas and thus also separate the growth rate into three individual growth rates $G_{n\mathrm{D}}$ ($n$ = 0, 1, 2). The illustrations for various flake sizes in Fig.~\ref{Fig:SupplyAreas}~a.i)-a.iv) clarify the relationship between the separate supply area sizes and flake sizes. In detail, the orange supply area remains constant (independent of spatial dimensions of the flake: 0D), the green supply area is proportional to the edge length of the flake (dependent on one spatial dimension: 1D), and the blue supply area corresponds to the flake area itself (dependent on two spatial dimensions: 2D). These three supply areas leading to the corresponding growth rates $G_{n\mathrm{D}}$ will be discussed individually in the following paragraphs.

At the very beginning, when no flake is present, only the flake size independent 0D supply area contributes to the growth or -- in this very particular case -- to the formation of the first 2D TMDC flake by its corresponding growth rate $G_\mathrm{0D}$. This case is visualized in Fig.~\ref{Fig:SupplyAreas}~a.i), in which only the circle-shaped orange supply area is present. The size of this area depends on the adsorption and desorption rates as well as on the diffusion constant of precursor material on the substrate and thus mirrors the probability of the event of randomly agglomerating precursor atoms adsorbed on the substrate surface.

Once a 2D TMDC flake is formed (corresponding to the blue triangle in Fig.~\ref{Fig:SupplyAreas}~a.ii)-a.iv)), the additional 1D and 2D supply areas emerge. The background of the 1D supply area (green) and its corresponding growth rate $G_\mathrm{1D}$ is, that precursor material adsorbs on the substrate surface near the flake, where it can diffuse to the flake and be built in at its edge (1D line of reactive sites) before desorbing. As the distance from which precursor material can diffuse to the flake edges is constant, the 1D supply area increases proportionally to the edge length $L$. Given that precursor material from this supply area is built in only at the edges of the flake, its contribution to the increase of the flake area $A$ is direct (direct growth). Therefore, this mechanism might be the most intuitive one. 

Because the third supply area (blue) is the flake area $A$ itself, in this case -- in contrast to the previously described 0D and 1D supply areas -- the precursor material is not adsorbed on the substrate, but on the already grown 2D TMDC flake, see Fig.~\ref{Fig:SupplyAreas}~b). It is very likely, that adsorption/desorption rates as well as diffusion constant differ for the precursor material on the flake itself from those on the substrate. If a flake is very small (Fig.~\ref{Fig:SupplyAreas}~a.ii)), material adsorbed on the flake is very likely to diffuse to the flake edge to contribute to the increase of the flake size by direct growth. The degree of material supply via the flake itself is expected to depend on the actual flake area $A$ for small flakes. Once a flake becomes rather large (Fig.~\ref{Fig:SupplyAreas}~a.iv)), the material supply via the flake and thus the growth would also become proportional to the edge length $L$ due to a limited diffusion range. We propose that the material supply via the flake area $A$ itself, even for very large flake areas, depends (approximately) on $A$, resulting in the growth rate $G_\mathrm{2D}$ within the 2D term of Eq.~(\ref{Eq:TotalDE}) valid for a wide range of flake sizes. This hypothesis is derived from the fact, that the precursor material consists of the same atoms as the 2D TMDC flake. Therefore, we anticipate complex, dynamic mechanisms taking place on/within the flake itself. The dynamics are discussed in more detail below, once the mechanisms underlying the etching rate are described.

For the etching rate $E$, we again introduce individual rates, $E_\mathrm{1D}$ (proportional to the flake edge length) and $E_\mathrm{2D}$ (proportional to the flake area). A flake size independent etching rate $E_\mathrm{0D}$ is not considered because no etching takes place without a flake being present. As soon as a few atoms agglomerate, already the smallest resulting agglomerate (nanoflake) has a spatial extension, so its decrease in size can be described by the size dependent etching rates $E_\mathrm{1D}$ and $E_\mathrm{2D}$.
 
The etching rate $E_\mathrm{1D}$ mirrors that 2D TMDC flakes have an increased chemical reactivity at their open edges with respect to their pristine basal planes. This manifests, for example, in an increased reactivity with oxygen \cite{Longo.2017} or in an increased catalytical activity at the edges \cite{Lauritsen.2003, Jaramillo.2007, Madau.2018}. Therefore, it seems intuitively reasonable, that etching also occurs preferentially at 2D TMDC edges as reported by Lv et al. for pristine 2D TMDC nanoflakes \cite{Lv.2017}. Obviously, the release of built-in atoms at edges (enhanced by oxygen) directly contributes to the decrease of the flake size.

Less intuitive are the mechanisms for the change of the flake size that may account for the etching rate $E_\mathrm{2D}$ as well as for the growth rate $G_\mathrm{2D}$. In the following, we will discuss various dynamically interacting mechanisms on and within the 2D TMDC flake at elevated temperatures. 

Firstly, we start with 2D TMDC flakes heated up to temperatures significantly lower than their typical CVD temperatures (>~650~\textdegree C) and at or close to ambient pressure. In this case, it is experimentally observed that the flakes begin to degrade and finally decompose completely \cite{Zhou.2013, Wu.2013, Yamamoto.2013, Ye.2016, Lv.2017, Maguire.2019, Yao.2020, Kaupmees.2020, Cullen.2021} [and see SI~1 (Fig.~S1)]. However, during this kind of degradation, the flakes do not become smaller from the edges. Instead, the atoms are released also from the basal planes of the 2D TMDC flakes. In some of these studies, annealing was intentionally performed with oxygen being present in the atmosphere. These include a comprehensive study by Cullen et al. showing for ten of the most common TMDCs degradation under ambient conditions \cite{Cullen.2021}. For all of these TMDCs the degradation temperature is spectroscopically determined to be (far) below 400~\textdegree C. That oxygen plays important role in the etching process is experimentally evident from the study by Yamamoto et al.: While no etching takes place in 2D MoS$_{2}$ under Ar/H$_{2}$ atmosphere at 350~\textdegree C, etching is observed under Ar/O$_{2}$ atmosphere already at temperatures around 300~\textdegree C \cite{Yamamoto.2013}. However, we observe such an etching effect in 2D WS$_{2}$ under completely inert Ar atmosphere at temperatures above 300~\textdegree C \cite{Kaupmees.2020} [and SI~1 (Fig.~S1)]. The reason might be small leakages that still let small amounts of air (oxygen) into the annealing system. On the other hand, because hydrogen binds oxygen, and because basal plane etching occurs in 2D MoS$_{2}$ even under Ar/H$_{2}$ atmospheres at temperatures in the range of 400-500~\textdegree C \cite{Ye.2016}, oxygen apparently only has a promotive role but is not necessary. If oxygen is present, the formation energy of S vacancies in pristine 2D MoS$_{2}$ basal planes becomes indeed negative \cite{Lin.2015}. However, the calculated oxygen dissociative adsorption barrier on pristine MoS$_{2}$ is rather large (1.59~eV) \cite{KC.2015}. At sites of S vacancies in the basal plane of 2D MoS$_{2}$, the oxygen dissociative adsorption barrier is halved \cite{KC.2015}. Hence, it is more likely to extend pre-existing defects (a certain number of intrinsic defects is always present) in the basal plane of 2D TMDCs, than to create new ones. This is supported by experiments giving evidence for grain boundaries and induced vacancies to be the preferred sites for the release of built-in atoms \cite{Lv.2017, Maguire.2019, Yao.2020, Kaupmees.2020} [and SI~1 (Fig.~S1)] and by studies showing the creation of defect clusters in form of triangular pits in 2D TMDCs \cite{Ye.2016, Zhou.2013, Wu.2013, Yamamoto.2013, Lv.2017, Maguire.2019, Kaupmees.2020}. The latter mechanism is often referred to as \textit{anisotropic (oxidative) etching}. The increased chemical reactivity of defect sites is consistent with experimental studies reporting a high catalytic activity of defect sites in 2D TMDCs \cite{Li.2016b, Yin.2016, Li.2016, Sun.2019}, rendering these sites to be chemically more like 2D TMDC edges than the pristine basal planes. Density-functional theory (DFT) calculations further confirm the increased chemical reactivity (catalytic activity as well as oxidation) at defect sites in the basal plane of 2D TMDC flakes \cite{Le.2014, KC.2015, Li.2016b, Li.2016, Dong.2018}. 

\begin{figure}[t]
\centering
\includegraphics[width=1.0\textwidth]{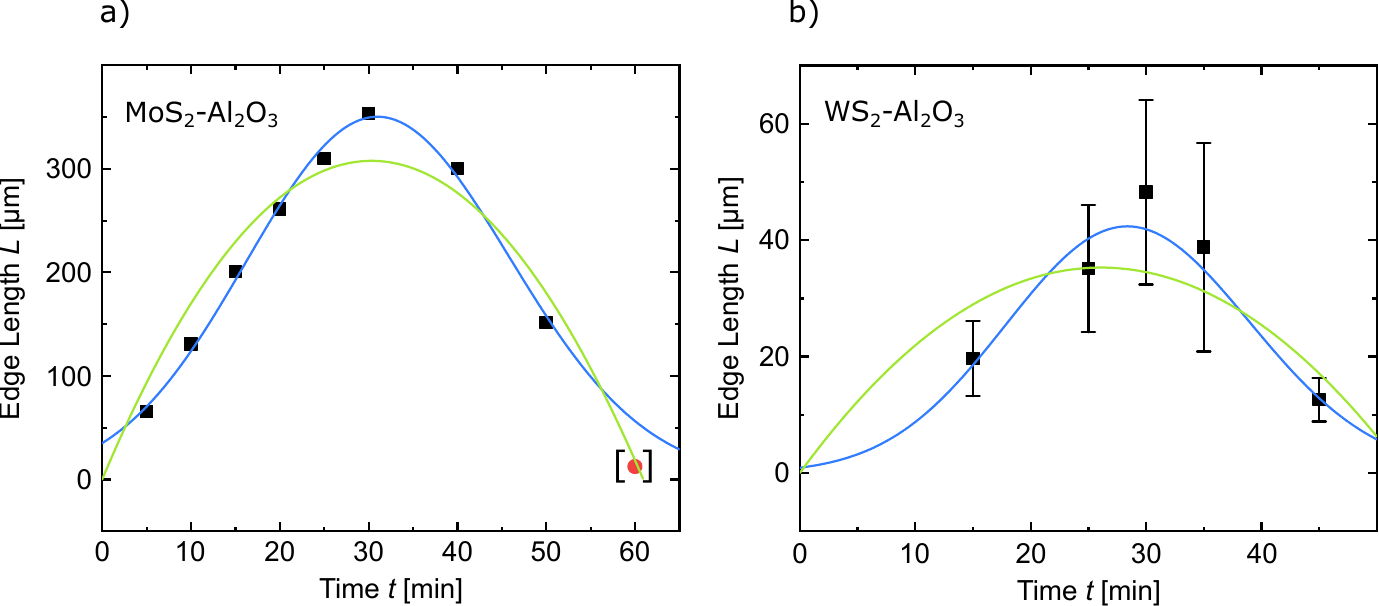}
\caption{Application of the 1D solution~(\ref{Eq:1D-Solution}) (green) and the 2D solution~(\ref{Eq:2D-Solution}) (blue) on experimental data a) of Chen et al. [Adapted with permission from \cite{Chen.2015}. Copyright 2015 American Chemical Society] b) of ourselves. The 2D solution can be fitted much better to both data sets. Red data point: Excluded because the growth rate is much lower than the etching rate resulting in fragmentation/anisotropic etching of the 2D TMDC flake.}
\label{Fig:ExperimentalData}
\end{figure}

From the previous paragraph we conclude, that etching does not only apply to the edges of 2D TMDC flakes ($L$-dependent/1D component), it also has an $A$-dependent/2D component ($E_\mathrm{2D}$). However, the etching process at the basal plane does not result in a reduction of the 2D TMDC flake size as reported by Chen et al. \cite{Chen.2015} (relevant data points in Fig.~\ref{Fig:ExperimentalData}~a)) and as shown by our own data in Fig.~\ref{Fig:ExperimentalData}~b). The major differences between this experimental observation of shrinking flakes and the studies mentioned in the previous paragraph are the conditions, under which the experiments are performed: (i) the latter are performed at much lower temperatures and (ii) in the absence of (at least one) precursor atom species.

If higher temperatures are applied, i.e. temperatures typically used in CVD and thus in the studies showing shrinking flakes (Fig.~\ref{Fig:ExperimentalData}), diffusion is also facilitated. For example, it has been shown by transmission electron microscopy (TEM), that even already built-in atoms are able to diffuse within the 2D TMDC lattice, if a neighbouring atomic site is empty (vacancy) \cite{Komsa.2013, Lin.2015, Hong.2017, Chen.2018, Maksov.2019}. This effect could be called also defect/vacancy diffusion. Once a vacancy reaches a 2D TMDC flake edge due to diffusion, the defect vanishes by reducing the flake size. This mechanism corresponds to an $A$-dependent etching component.

The vacancy diffusion barrier has been calculated by DFT for 2D MoS$_{2}$ and MoSe$_{2}$ to be between 0.6 and 2.9~eV \cite{Komsa.2013, Le.2014, Lin.2015, Yu.2015, Sensoy.2017, Hong.2017, Chen.2018}. The actual calculated barrier value depends on the type of vacancy (transition metal vacancy, single chalcogen vacancy, double chalcogen vacancy) and on the environment of the diffusing vacancy. For instance, the diffusion barrier of a single S vacancy is strongly reduced once an additional vacancy exists on a neighbouring site \cite{Komsa.2013, Le.2014}. Therefore, pairs of single S vacancies would migrate faster through the 2D TMDC lattice. Because existing single S vacancies facilitate oxidation \cite{KC.2015} as discussed above and thus improve the release of neighbouring S atoms, the basal plane etching may initially enhance the vacancy diffusion velocity. When S vacancies agglomerate, they tend to form vacancy rows as experimentally observed by TEM even at room temperature \cite{Komsa.2013, Chen.2018}. At high temperatures, the number of rows decreases, but their length increases \cite{Chen.2018}. DFT calculations confirm that these vacancy rows are energetically favored \cite{Komsa.2013, Le.2014}. Due to a large diffusion barrier, S atoms at the edge of the vacancy row are unlikely to diffuse into the row \cite{Le.2014}. Instead, atoms within the vacancy rows (both S and Mo atoms for 2D MoS$_{2}$) are able to migrate rapidly through the lattice \cite{Chen.2018}. Because in this way a lot of material can be moved, such "channels" are important for the formation of triangular pits within the basal planes of 2D TMDCs \cite{Chen.2018}. In this way, at 800~\textdegree C, a triangular pit with a diameter of a few nm can be opened at the end of a vacancy row within one minute.  

It is very likely, that the vacancy diffusion observed experimentally is triggered by the high kinetic energy of the electrons during the TEM measurements mostly performed at room temperature. However, Lin et al. report similar morphological structures formed by defect diffusion within a 2D MoS$_{2}$ lattice after annealing at 700~\textdegree C in high vacuum as within the 2D MoSe$_{2}$ lattice after extensive defect diffusion triggered due to the electron beam \cite{Lin.2015}. Hence, such high temperatures, which are also typical for CVD of 2D TMDCs, may also be sufficient for a reasonable high diffusion of defects.

In addition, during CVD, simultaneous to defect creation and diffusion, the growth still takes place, i.e. new precursor atoms adsorb, diffuse, and are built in, if they reach an appropriate site. Not only edges are appropriate sites, but also the diffusing vacancies. Hence, adsorbed precursor atoms, which statistically would not be able to reach the edge (diffusion range), would at least compensate the reduction of the flake size due to the release of built-in atoms and subsequent defect diffusion to the edges (indirect growth). In SI~1 Fig.~S1 we demonstrate the influence of precursor atoms existing in the gas phase on the degradation velocity. We compare the degradation of 2D WS$_{2}$ flakes under Ar, and sulphur containing Ar atmosphere and find a reduced degradation velocity, if S atoms are present.

Summarising up to here, a CVD process is not only about the growth of the flakes. Rather, CVD is a highly dynamic process including adsorption, etching, diffusion (of adsorbed atoms and vacancies), agglomeration, healing, and growth. In this respect, the most dynamic region during growth is the flake surface $A$ itself.

Taking all contributions to growth and etching into account, we arrive at the following differential equation to describe the dynamic behaviour of a 2D TMDC flake during synthesis:

\begin{equation}
\frac{\mathrm{d}A}{\mathrm{d}t} = [G_\mathrm{2D}(t) - E_\mathrm{2D}]\, A + [G_\mathrm{1D}(t) - E_\mathrm{1D}]\, L + G_\mathrm{0D}(t)
\label{Eq:TotalDE}
\end{equation}

This is the basic equation of our dynamic growth/etch model. In general, and in accordance with the previous consideration of the dynamic mechanisms during CVD, this equation is composed of three terms with different dependencies on the flake size: one 2D term and one 1D term reflecting either the dependence of the changing flake size on the flake area $A$ or the edge length $L$, and one flake size-independent 0D term.

Actually, besides the growth rates $G_{n\mathrm{D}}$, also the etching rates $E_{n\mathrm{D}}$ in Eq.~(\ref{Eq:TotalDE}) would be time-dependent. While for the latter the time dependence stems from their dependencies on temperature $T(t)$ and pressure $p(t)$, the growth rates additionally depend on the amount of precursor source material $M(t)$. For our following discussion and application of Eq.~(\ref{Eq:TotalDE}), we assume a constant temperature and a constant pressure during the entire growth process with the duration $t$. Therefore, only the growth rates $G_{n\mathrm{D}}$ depend on $t$, or more precisely, on $M(t)$.

Further, we reduce our model to only one solid precursor source resulting in the schematic process configuration shown in Fig.~\ref{Fig:SupplyAreas}~c). The mathematical description of the growth rates $G_{n\mathrm{D}}$ for only one solid precursor source, which is introduced below in the solution~(\ref{Eq:Solution_GrowthRates}), are a good approximation for many cases under following conditions: either, when most of the time one of the two precursor atom species is abundant relative to the other one, and/or, when the specific evaporation rates $v$ of both precursor sources are approximately equal (see SI~2 for more details). The reduction to one solid precursor source is also experimentally confirmed by the typically (equilateral) triangular shape of 2D TMDC flakes grown by CVD \cite{Ye.2016, Li.2016, Madau.2018, Lee.2013, Zhang.2013, Huang.2014, Ling.2014, Wang.2014, Okada.2014, Wang.2014b, Ago.2015, Dumcenco.2015, Liu.2016, Chen.2015, Cain.2016, Zhu.2017, Zhang.2019, Pollmann.2020, Okada.2020, Pollmann.2020b, Zhang.2019b, Lv.2017, Kaupmees.2020} as this type of shape occurs, when one of the two precursor atom species is abundant relative to the other one \cite{Wang.2014b, Zhu.2017}. Therefore, it is justified to take only one solid precursor source into account for the reaction rate and thus the growth rates $G_{n\mathrm{D}}$ in CVD.

From the typically (equilateral) triangular shape of grown 2D TMDC flakes, we derive the relationship $A = \frac{\sqrt{3}}{4} L^{2}$. Hence, our basic equation in principle can be solved. Furthermore, as the flake size is often expressed by the edge length $L$ (or by the lateral size in one spatial dimension) in literature \cite{Ye.2016, Li.2016, Zhang.2013, Huang.2014, Ling.2014, Wang.2014, Okada.2014, Wang.2014b, Ago.2015, Chen.2015b, Chen.2015, Pollmann.2020, Okada.2020, Lv.2017, Kaupmees.2020}, the solutions and results are presented as a function of the edge length $L$ below. Nevertheless, we decided to use $\frac{\mathrm{d}A}{\mathrm{d}t}$ in Eq.~(\ref{Eq:TotalDE}) because $A$ is directly proportional to the mass of the flake $m$ using the two-dimensional density of one TMDC layer $\rho_\mathrm{2D}$ and, thus, also to the number of built-in atoms. Therefore, $\frac{\mathrm{d}A}{\mathrm{d}t}$ is proportional to the mass change $\frac{\mathrm{d}m}{\mathrm{d}t}$. We believe that this convention renders our basic differential equation~(\ref{Eq:TotalDE}) to be more intuitively understandable. 

Next, we want to comment on the $t$-dependency of the growth rates $G_{n\mathrm{D}}$. As this $t$-dependency stems from the depletion of the solid precursor source during a running CVD process -- either by being consumed or by forming a passivation layer on its surface (so-called poisoning) \cite{Chen.2015} --, we describe the growth rates by the differential equation

\begin{equation}
G_{n\mathrm{D}}(t) = - c_{n\mathrm{D}} \frac{\mathrm{d}M}{\mathrm{d}t} = c_{n\mathrm{D}} v M(t),
\label{Eq:DE_GrowthRates}
\end{equation}

\noindent with the solution (initial condition: $M (t = 0) = M_{0}$)

\begin{equation}
G_{n\mathrm{D}}(t) = c_{n\mathrm{D}} v M_{0}\, \mathrm{e}^{-vt}.
\label{Eq:Solution_GrowthRates}
\end{equation}

Equation~(\ref{Eq:DE_GrowthRates}) expresses that the growth rates $G_{n\mathrm{D}}$ ($n$ = 0, 1, 2) are proportional to the temporal change of the precursor source mass $-\frac{\mathrm{d}M}{\mathrm{d}t}$. In other words, the more material from the material source moves into the gas phase, the more material can adsorb on the substrate surface (including the surface of the already grown 2D TMDC flakes) and contribute to the growth of 2D TMDC flakes. On the other hand, at constant temperatures, $-\frac{dM}{dt}$ is also proportional to the mass of material available in the precursor source $M$. $v$ is the specific vaporisation rate and $c_{n\mathrm{D}}$ is a proportionality factor which takes into account microscopic as well as macroscopic factors. The latter include transport losses (see $c_\mathrm{macros.}$ in Fig.~\ref{Fig:SupplyAreas}~c)) and the fact, that one precursor source supplies hundreds of flakes simultaneously. The microscopic factors are different for the three supply areas because of varying conditions for e.g. adsorption, desorption, diffusion and reaction probabilities (see $c_\mathrm{micros.}$ in Fig.~\ref{Fig:SupplyAreas}~b)). Hence, $c_{n\mathrm{D}}$ in general is specific for each of the terms in Eq.~(\ref{Eq:TotalDE}).

Solution~(\ref{Eq:Solution_GrowthRates}) for $G_{n\mathrm{D}}(t)$ is a strictly monotonically decreasing function. Hence, with adequate constants $c_{n\mathrm{D}}$ and $M_{0}$, the growth rates $G_{n\mathrm{D}}$ initially dominate over the etching rates $E_{n\mathrm{D}}$ in Eq.~(\ref{Eq:TotalDE}). The specific vaporisation rate $v$ finally leads to a dominance of the etching rate $E$. Therefore, solutions of Eq.~(\ref{Eq:TotalDE}) can describe 2D TMDC flakes, which first become larger and later on shrink again with time, and can thus in principle explain the experimental observations of Chen et al. \cite{Chen.2015} and our own data in Fig.~\ref{Fig:ExperimentalData}.

\begin{figure}[t]
\centering
\includegraphics[width=1.0\textwidth]{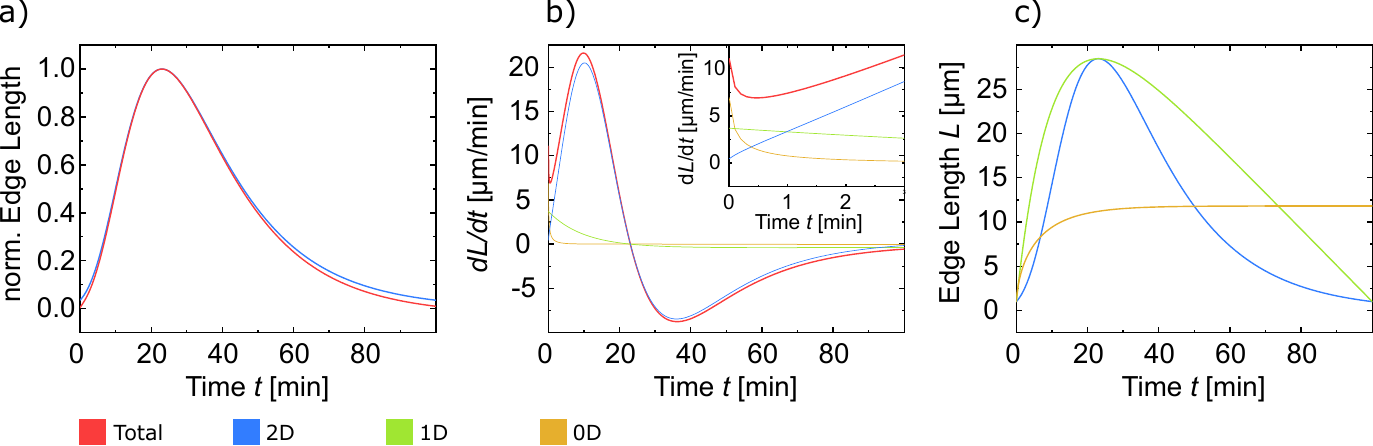}
\caption{Exemplary solutions of the total differential equation~(\ref{Eq:TotalDE}) and of the partial differential equations~(\ref{Eq:0D-Eq})-(\ref{Eq:2D-Eq}). a) Numerical solution of the total differential equation and the 2D solution~\ref{Eq:2D-Solution}. b) Derivation of the total solution and the contribution of the three terms to the total differential equation~\ref{Eq:TotalDE}. c) Analytical 0D, 1D, and 2D solutions~(\ref{Eq:0D-Solution})-(\ref{Eq:2D-Solution}).}
\label{Fig:DE-Solutions}
\end{figure}

Next, we will present and discuss actual solutions of the total as well as parts of the differential equation~(\ref{Eq:TotalDE}). Unfortunately, the total differential equation~(\ref{Eq:TotalDE}) is not analytically solvable with the growth rates from Eq.~(\ref{Eq:Solution_GrowthRates}). We therefore begin with its numerical solution (see methods). The typical shape, namely, first growth rates $G$ and then the etching rates $E$ become dominant, is shown in Fig.~\ref{Fig:DE-Solutions}~a), red curve. Initially, the curve rapidly increases to a maximum of the flake size (here: flake edge length $L$). Thereafter, it drops somewhat less rapidly, but still rather quickly. 

Figure~\ref{Fig:DE-Solutions}~b) shows the derivative of the numerical solution (red) including the contributions of the three terms of Eq.~(\ref{Eq:TotalDE}). Obviously, after a short time, the 2D term (blue) becomes predominant in growth and etching. The inset of Fig.~\ref{Fig:DE-Solutions}~b) illustrates that the 0D term (orange) and then the 1D term (green) dominate in the early stages of the flake growth.

In order to identify the leading term(s) and thus the dominant physical mechanism(s) we split the total equation into the following three partial equations, each for one of term:

\begin{subequations}
\begin{align}
\frac{\mathrm{d}A}{\mathrm{d}t} &= G_\mathrm{0D}(t)
\label{Eq:0D-Eq} \\
\frac{\mathrm{d}A}{\mathrm{d}t} &= [G_\mathrm{1D}(t) - E_\mathrm{1D}]\, L
\label{Eq:1D-Eq} \\
\frac{\mathrm{d}A}{\mathrm{d}t} &= [G_\mathrm{2D}(t) - E_\mathrm{2D}]\, A
\label{Eq:2D-Eq}
\end{align}
\end{subequations}

The analytical solutions -- again in terms of the solution for the respective growth rate [solution~(\ref{Eq:Solution_GrowthRates})] -- are as follows (with $L(t=0) = L_{0}$):

\begin{subequations}
\begin{align}
L_\mathrm{0D}(t) &= \left[L_{0}^{2} + \frac{4c_\mathrm{0D}M_{0}\, (1-\mathrm{e}^{-vt})}{\sqrt{3}}\right]^{\frac{1}{2}}
\label{Eq:0D-Solution} \\
L_\mathrm{1D}(t) &= L_{0} + \frac{2}{\sqrt{3}} \left[-E_\mathrm{1D}\, t + c_\mathrm{1D}M_{0}\, (1-\mathrm{e}^{-vt})\right]
\label{Eq:1D-Solution} \\
L_\mathrm{2D}(t) &= L_{0} \, \exp\left[\frac{-E_\mathrm{2D}\, t + c_\mathrm{2D}M_{0}\, (1-\mathrm{e}^{-vt})}{2}\right]
\label{Eq:2D-Solution}
\end{align}
\end{subequations}

Note, that a closed form solution also exists for the 1D2D differential equation, i.e. equation~\ref{Eq:TotalDE} without the 0D term. This solution is in fact to complex to be of practical value. However, we show and discuss it in SI 3.

Exemplary quantitative curves of these solutions are shown in Fig.~\ref{Fig:DE-Solutions}~c) and their parameters are listed in SI~4 (Tab.~S2). The parameters were chosen so that for the 1D solution~(\ref{Eq:1D-Solution}) and the 2D solution~(\ref{Eq:2D-Solution}) the local maxima are congruent. Without an etching rate, the 0D solution~(\ref{Eq:0D-Solution}) has no local maximum of course, but it is a monotonically increasing function with the largest change existing for $t \rightarrow 0$. The solution of the 1D and 2D partial equations differ in such a way that the curve of the 1D solution is more convex near the local maximum, while the curve of the 2D solution is comparatively sharp. Mathematically, this behaviour is related to the nature of the two solutions: the 2D solution corresponds in its form to the exponential function of the 1D solution.

It becomes evident that the solution of the 2D partial equation~(\ref{Eq:2D-Solution}) is very similar in shape to the numerical solution of the total differential equation~(\ref{Eq:TotalDE}). In order to elucidate this, the 2D solution normalized to its maximum has been added to the plot of the numerical solution in Fig.~\ref{Fig:DE-Solutions}~a). The 2D solution mainly diverges from the total solution for small or large times (both corresponding to small flakes), which is consistent with the expectation from Fig.~\ref{Fig:DE-Solutions}~b) that the total differential equation~(\ref{Eq:TotalDE}) and its solution is largely dominated by its 2D term. Hence, the 2D solution~(\ref{Eq:2D-Solution}) is widely applicable as an analytically derived approximation for the actual solution of the total differential equation. In this context, the parameter $L_{0}$ must be chosen in order to compensate the neglected nucleation and early growth stages.

In order to test our model for plausibility, we apply it to experimental data of Chen et al. for 2D MoS$_{2}$ grown on sapphire \cite{Chen.2015} and on our own data for 2D WS$_{2}$ grown on sapphire. Both data sets are acquired by analysing flake size distributions from several growth processes for a varying growth duration $t$ at maximum (growth) temperature and are shown in Fig.~\ref{Fig:ExperimentalData}~a) and Fig.~\ref{Fig:ExperimentalData}~b), respectively. Because the 2D solution~(\ref{Eq:2D-Solution}) (blue) can be fitted to both data sets obviously better than the 1D solution~(\ref{Eq:1D-Solution}) (green), our model and in particular our hypothesis, that the 2D term is largely predominant for the 2D TMDC growth, is confirmed.

Now, we will discuss the data point at 60~min in the data set of Chen et al. (red in brackets in Fig.~\ref{Fig:ExperimentalData}~a)). Both, the 1D and 2D solutions are fitted to the data set except for this last data point. However, even if it is included, this data point significantly differs from the best fit of the 2D solution and, thus, increases the fitting error, see SI~5.

Considering the history of this data point from Chen et al. \cite{Chen.2015}, it is possible not only to justify why the data point can be excluded in Fig.~\ref{Fig:ExperimentalData}~a), but it even confirms the premises for our model. For all other data points, our model of initially growing and later shrinking flakes can be applied. At the growth duration of 60 min, in fact, Chen et al. found that the grown 2D MoS$_{2}$ flakes behave differently. At this time anisotropic oxidative etching is observed: Instead of shrinking, triangular pits are formed within the basal plane of the flake resulting in a fragmentation into many smaller triangular flakes. This fragmentation/anisotropic etching effect is beyond our model, which does not take into account degradation of the basal plane occurring once the growth rate is significantly smaller than the etching rate. Apparently, Chen et al. evaluated the size of the small flake fragments for their data set, so the data point is shifted towards lower values as expected from our fit of the 2D solution.

Moreover, the fragmentation of the flakes proves that etching occurs not only at the edges, but also and in particular on the basal plane of the flakes. Obviously, this etching mechanism is only observable at later stages of the growth process, once the supply of precursor atoms is significantly diminished. The reduced quantity of new atoms can no longer sufficiently compensate the etching of the basal plane, resulting in the formation of defect clusters turning into pits with a low mobility that prevent the diffusion to the edges. Subsequently, the pits continue to enlarge anisotropically in the preferred directions of the TMDC lattice, becoming triangular, which is typically found for the flake degradation in absence of precursor atoms \cite{Ye.2016, Zhou.2013, Wu.2013, Yamamoto.2013, Lv.2017, Maguire.2019, Kaupmees.2020, Chen.2018}. 

\section{Conclusion}

In summary, we have presented a mathematical concept to describe the growth as well as the etching processes of 2D TMDCs during their synthesis at high temperatures. The individual mechanisms are represented by a differential equation that is made up of three parts. By a detailed analysis we found: (i) the numerical solution of the total differential equation differs only very slightly from the 2D solution in wide range, (ii) the growth and shrinking phenomenon experimentally found can be best analytically approximated by the 2D solution. These results imply, that both the material supply for growth and the material loss/etching are largely determined by the size of the flake area. This is in contrast to the common assumption of the importance and dominance of the flake edges, also suggested (but not discussed) by Chen et al. Our findings are corroborated by the fact, that the 1D solution corresponds to growth dominated by edges and the fits to both experimental data sets, including the one of Chen et al., are rather poor.

The predominance of the 2D solution of our model alters and advances the understanding of how the synthesis of 2D TMDCs takes place. It is based on the interplay between highly dynamic mechanisms at the atomic level within the flake itself and is thus consistent with the expectations at the high temperatures typically used for synthesis.

Our model provides an explanation for the rapid growth of 2D TMDC flakes by CVD (typical average: order of 100~nm/s). In order to exceed the etching term for a sufficient period of time in solid precursor source based CVD to actually obtain 2D TMDCs as product, the initial growth rates have to be chosen unphysically large. This clearly reveals that metalorganic CVD has the key advantage of a continuous supply of material over time. Within our model, the growth rate of metalorganic CVD would no longer be time dependent because of non-depleting precursor sources. This allows the etching rate to be precisely compensated and a constant net growth rate to be set. Note however, that the method suffers from other disadvantages such as small grain sizes, for example Kang et al. \cite{Kang.2015}.

Because the actual synthesis of TMDC monolayer is based on the in- and on-flake dynamics due to the high process temperature, we finally conclude with a hypothesis for the frequently observed mutlilayer growth. Because the dynamic processes decay with decreasing temperature, we believe, that multilayer growth primarily takes place in the cooling phase. That is, once the temperature is sufficiently reduced, a transition state with a time constant depending on the cooling rate appears. Within this transition, remaining precursor atoms would continue to adsorb on the flake surface. Because their diffusion constant as well as the release of atoms (i.e. the etching rate is reduced at low temperatures, the precursor atoms might merge to additional layers on the first one. Optimizing growth during the cooling phase could thus be a successful strategy to suppress or enhance bilayer formation.

\section{Materials and Methods}

\paragraph{Chemical Vapour Deposition}

Tungsten disulphide (WS$_{2}$) flakes for the time-de\-pen\-dent study shown in Fig.~\ref{Fig:ExperimentalData}~b) has been grown by chemical vapour deposition on c-face sapphire substrates. Therefore a custom made process system consisting of a heating belt and a tube furnace (ThermConcept ROS 38/250/12) was used, that in this way provides two heating zones in a quartz tube. A mass flow controller is used to adjust an argon (Air Liquide, 99.999~\%) flow through the tube. The sapphire substrates are cleaned by an ultra sonic bath in ethanol and prepared by homogeneously spreading of individual WO$_{3}$ powder grains (Alfa Aesar, 99.8~\%) on their surfaces. The substrates are deposited in a ceramic crucible in the downstream heating zone (tube furnace). In the upstream heating zone (heating belt) a second crucible with 160~mg sulphur (Sigma-Aldrich, 99.98~\%) is deposited. After sealing the quartz tube, it is flushed by argon gas. The upstream heating zone is heated up to 150~\textdegree C and the downstream heating zone to 800~\textdegree C. The maximum temperatures are hold for 15 to 45~min. During the whole process a constant argon flow of 10~Ncm$^{3}$/min is used. The pressure in the tube was close to ambient pressure.

\paragraph{Numerical Solving}

Differential Equation~\ref{Eq:TotalDE} is solved numerically by Mathematica (Wolfram Research, Inc.). 

\section*{Acknowledgement}

The authors acknowledge the German Research Foundation (DFG) by funding SCHL 384/20-1 (project number 406129719).

\bibliography{Bib/Bib}

\begin{thebibliography}{10}

\bibitem{Radisavljevic.2011}
B.~Radisavljevic, A.~Radenovic, J.~Brivio, V.~Giacometti, and A.~Kis.
\newblock {Single-Layer MoS$_{2}$ Transistors}.
\newblock {\em {Nat. Nanotechnol.}}, 6(3):147--150, 2011.

\bibitem{Yin.2012}
Z.~Yin, H.~Li, H.~Li, L.~Jiang, Y.~Shi, Y.~Sun, G.~Lu, Q.~Zhang, X.~Chen, and
  H.~Zhang.
\newblock {Single-Layer MoS$_{2}$ Phototransistors}.
\newblock {\em {ACS Nano}}, 6(1):74--80, 2012.

\bibitem{LopezSanchez.2013}
O.~Lopez-Sanchez, D.~Lembke, M.~Kayci, A.~Radenovic, and A.~Kis.
\newblock {Ultrasensitive Photodetectors Based on Monolayer MoS$_{2}$}.
\newblock {\em {Nat. Nanotechnol.}}, 8(7):497--501, 2013.

\bibitem{Li.2012}
H.~Li, Z.~Yin, Q.~He, H.~Li, X.~Huang, G.~Lu, D.~W.~H. Fam, A.~I.~Y. Tok,
  Q.~Zhang, and H.~Zhang.
\newblock {Fabrication of Single- and Multilayer MoS$_{2}$ Film-Based
  Field-Effect Transistors for Sensing NO at Room Temperature}.
\newblock {\em {Small}}, 8(1):63--67, 2012.

\bibitem{Urban.2019}
F.~Urban, F.~Giubileo, A.~Grillo, L.~Iemmo, G.~Luongo, M.~Passacantando,
  T.~Foller, L.~Madau{\ss}, E.~Pollmann, M.~P. Geller, D.~Oing, M.~Schleberger,
  and A.~{Di Bartolomeo}.
\newblock {Gas Dependent Hysteresis in MoS$_{2}$ Field Effect Transistors}.
\newblock {\em {2D Mater.}}, 6(4):045049, 2019.

\bibitem{Lauritsen.2003}
J.~V. Lauritsen, M.~Nyberg, R.~T. Vang, M.~V. Bollinger, B.~S. Clausen,
  {Tops{\o}e H.}, K.~W. Jacobsen, E.~L{\ae}gsgaard, J.~K. N{\o}rskov, and
  F.~Besenbacher.
\newblock {Chemistry of One-Dimensional Metallic Edge States in MoS$_{2}$
  Nanoclusters}.
\newblock {\em {Nanotechnology}}, 14(3):385--389, 2003.

\bibitem{Jaramillo.2007}
T.~F. Jaramillo, K.~P. J{\o}rgensen, J.~Bonde, J.~H. Nielsen, S.~Horch, and
  I.~Chorkendorff.
\newblock {Identification of Active Edge Sites for Electrochemical H$_{2}$
  Evolution from MoS$_{2}$ Nanocatalysts}.
\newblock {\em {Science}}, 317(5834):100--102, 2007.

\bibitem{Le.2014}
D.~Le, T.~B. Rawal, and T.~S. Rahman.
\newblock {Single-Layer MoS$_{2}$ with Sulfur Vacancies: Structure and
  Catalytic Application}.
\newblock {\em {J. Phys. Chem. C}}, 118(10):5346--5351, 2014.

\bibitem{Li.2016b}
H.~Li, C.~Tsai, A.~L. Koh, L.~Cai, A.~W. Contryman, A.~H. Fragapane, J.~Zhao,
  H.~S. Han, H.~C. Manoharan, F.~Abild-Pedersen, J.~K. N{\o}rskov, and
  X.~Zheng.
\newblock {Activating and Optimizing MoS$_{2}$ Basal Planes for Hydrogen
  Evolution through the Formation of Strained Sulphur Vacancies}.
\newblock {\em {Nat. Mater.}}, 15(1):48--53, 2016.

\bibitem{Ye.2016}
G.~Ye, Y.~Gong, J.~Lin, B.~Li, Y.~He, S.~T. Pantelides, W.~Zhou, R.~Vajtai, and
  P.~M. Ajayan.
\newblock {Defects Engineered Monolayer MoS$_{2}$ for Improved Hydrogen
  Evolution Reaction}.
\newblock {\em {Nano Lett.}}, 16(2):1097--1103, 2016.

\bibitem{Yin.2016}
Y.~Yin, J.~Han, Y.~Zhang, X.~Zhang, P.~Xu, Q.~Yuan, L.~Samad, X.~Wang, Y.~Wang,
  Z.~Zhang, P.~Zhang, X.~Cao, B.~Song, and S.~Jin.
\newblock {Contributions of Phase, Sulfur Vacancies, and Edges to the Hydrogen
  Evolution Reaction Catalytic Activity of Porous Molybdenum Disulfide
  Nanosheets}.
\newblock {\em {J. Am. Chem. Soc.}}, 138(25):7965--7972, 2016.

\bibitem{Li.2016}
G.~Li, {Du Z.}, Q.~Qiao, Y.~Yu, D.~Peterson, A.~Zafar, R.~Kumar, S.~Curtarolo,
  F.~Hunte, S.~Shannon, Y.~Zhu, W.~Yang, and L.~Cao.
\newblock {All The Catalytic Active Sites of MoS$_{2}$ for Hydrogen Evolution}.
\newblock {\em {J. Am. Chem. Soc.}}, 138(51):16632--16638, 2016.

\bibitem{Dong.2018}
S.~Dong and Z.~Wang.
\newblock {Grain Boundaries Trigger Basal Plane Catalytic Activity for the
  Hydrogen Evolution Reaction in Monolayer MoS$_{2}$}.
\newblock {\em {Electrocatalysis}}, 9(6):744--751, 2018.

\bibitem{Madau.2018}
L.~Madau{\ss}, I.~Zegkinoglou, H.~{V{\'a}zquez Mui{\~n}os}, Y.-W. Choi,
  S.~Kunze, M.-Q. Zhao, C.~H. Naylor, P.~Ernst, E.~Pollmann, O.~Ochedowski,
  H.~Lebius, A.~Benyagoub, B.~Ban-d'Etat, A.~T.~C. Johnson, F.~Djurabekova,
  B.~{Roldan Cuenya}, and M.~Schleberger.
\newblock {Highly Active Single-Layer MoS$_{2}$ Catalysts Synthesized by Swift
  Heavy Ion Irradiation}.
\newblock {\em {Nanoscale}}, 10(48):22908--22916, 2018.

\bibitem{Sun.2019}
C.~Sun, P.~Wang, H.~Wang, C.~Xu, J.~Zhu, Y.~Liang, Y.~Su, Y.~Jiang, W.~Wu,
  E.~Fu, and G.~Zou.
\newblock {Defect Engineering of Molybdenum Disulfide through Ion Irradiation
  to Boost Hydrogen Evolution Reaction Performance}.
\newblock {\em {Nano Res.}}, 12(7):1613--1618, 2019.

\bibitem{Magda.2015}
G.~Z. Magda, J.~Pető, G.~Dobrik, C.~Hwang, L.~P. Bir{\'o}, and
  L.~Tapaszt{\'o}.
\newblock {Exfoliation of Large-Area Transition Metal Chalcogenide Single
  Layers}.
\newblock {\em {Sci. Rep.}}, 5:14714, 2015.

\bibitem{Velicky.2018}
M.~Velick{\'y}, G.~E. Donnelly, W.~R. Hendren, S.~McFarland, D.~Scullion,
  W.~J.~I. DeBenedetti, G.~C. Correa, Y.~Han, A.~J. Wain, M.~A. Hines, D.~A.
  Muller, K.~S. Novoselov, H.~D. Abru{\~n}a, R.~M. Bowman, E.~J.~G. Santos, and
  F.~Huang.
\newblock {Mechanism of Gold-Assisted Exfoliation of Centimeter-Sized
  Transition-Metal Dichalcogenide Monolayers}.
\newblock {\em {ACS Nano}}, 12(10):10463--10472, 2018.

\bibitem{Pollmann.2021}
E.~Pollmann, S.~Sleziona, T.~Foller, U.~Hagemann, C.~Gorynski, O.~Petri,
  L.~Madau{\ss}, L.~Breuer, and M.~Schleberger.
\newblock {Large-Area, Two-Dimensional MoS$_{2}$ Exfoliated on Gold: Direct
  Experimental Access to the Metal-Semiconductor Interface}.
\newblock {\em {ACS Omega}}, 6(24):15929--15939, 2021.

\bibitem{Lee.2012}
Y.-H. Lee, X.-Q. Zhang, W.~Zhang, M.-T. Chang, C.-T. Lin, K.-D. Chang, Y.-C.
  Yu, J.~T.-W. Wang, C.-S. Chang, L.-J. Li, and T.-W. Lin.
\newblock {Synthesis of Large-Area MoS$_{2}$ Atomic Layers with Chemical Vapor
  Deposition}.
\newblock {\em {Adv. Mater.}}, 24(17):2320--2325, 2012.

\bibitem{Lee.2013}
Y.-H. Lee, L.~Yu, H.~Wang, W.~Fang, X.~Ling, Y.~Shi, C.-T. Lin, J.-K. Huang,
  M.-T. Chang, C.-S. Chang, M.~Dresselhaus, T.~Palacios, L.-J. Li, and J.~Kong.
\newblock {Synthesis and Transfer of Single-Layer Transition Metal Disulfides
  on Diverse Surfaces}.
\newblock {\em {Nano Lett.}}, 13(4):1852--1857, 2013.

\bibitem{Yu.2013}
Y.~Yu, C.~Li, Y.~Liu, L.~Su, Y.~Zhang, and L.~Cao.
\newblock {Controlled Scalable Synthesis of Uniform, High-Quality Monolayer and
  Few-Layer MoS$_{2}$ Films}.
\newblock {\em {Sci. Rep.}}, 3:1866, 2013.

\bibitem{Zhang.2013}
Y.~Zhang, Y.~Zhang, Q.~Ji, J.~Ju, H.~Yuan, J.~Shi, T.~Gao, D.~Ma, M.~Liu,
  Y.~Chen, X.~Song, H.~Y. Hwang, Y.~Cui, and Z.~Liu.
\newblock {Controlled Growth of High-Quality Monolayer WS$_{2}$ Layers on
  Sapphire and Imaging its Grain Boundary}.
\newblock {\em {ACS Nano}}, 7(10):8963--8971, 2013.

\bibitem{Huang.2014}
J.-K. Huang, J.~Pu, C.-L. Hsu, M.-H. Chiu, Z.-Y. Juang, Y.-H. Chang, W.-H.
  Chang, Y.~Iwasa, T.~Takenobu, and L.-J. Li.
\newblock {Large-Area Synthesis of Highly Crystalline WSe$_{2}$ Monolayers and
  Device Applications}.
\newblock {\em {ACS Nano}}, 8(1):923--930, 2014.

\bibitem{Ling.2014}
X.~Ling, Y.-H. Lee, Y.~Lin, W.~Fang, L.~Yu, M.~S. Dresselhaus, and J.~Kong.
\newblock {Role of the Seeding Promoter in MoS$_{2}$ Growth by Chemical Vapor
  Deposition}.
\newblock {\em {Nano Lett.}}, 14(2):464--472, 2014.

\bibitem{Wang.2014}
X.~Wang, Y.~Gong, G.~Shi, W.~L. Chow, K.~Keyshar, G.~Ye, R.~Vajtai, J.~Lou,
  Z.~Liu, E.~Ringe, B.~K. Tay, and P.~M. Ajayan.
\newblock {Chemical Vapor Deposition Growth of Crystalline Monolayer
  MoSe$_{2}$}.
\newblock {\em {ACS Nano}}, 8(5):5125--5131, 2014.

\bibitem{Okada.2014}
M.~Okada, T.~Sawazaki, K.~Watanabe, T.~Taniguch, H.~Hibino, H.~Shinohara, and
  R.~Kitaura.
\newblock {Direct Chemical Vapor Deposition Growth of WS$_{2}$ Atomic Layers on
  Hexagonal Boron Nitride}.
\newblock {\em {ACS Nano}}, 8(8):8273--8277, 2014.

\bibitem{Wang.2014b}
S.~Wang, Y.~Rong, Y.~Fan, M.~Pacios, H.~Bhaskaran, K.~He, and J.~H. Warner.
\newblock {Shape Evolution of Monolayer MoS$_{2}$ Crystals Grown by Chemical
  Vapor Deposition}.
\newblock {\em {Chem. Mater.}}, 26(22):6371--6379, 2014.

\bibitem{Ago.2015}
H.~Ago, H.~Endo, P.~Sol{\'i}s-Fern{\'a}ndez, R.~Takizawa, Y.~Ohta, Y.~Fujita,
  K.~Yamamoto, and M.~Tsuji.
\newblock {Controlled van der Waals Epitaxy of Monolayer MoS$_{2}$ Triangular
  Domains on Graphene}.
\newblock {\em {ACS Appl. Mater. Interfaces}}, 7(9):5265--5273, 2015.

\bibitem{Dumcenco.2015}
D.~Dumcenco, D.~Ovchinnikov, K.~Marinov, P.~Lazi{\'c}, M.~Gibertini,
  N.~Marzari, O.~{Lopez Sanchez}, Y.-C. Kung, D.~Krasnozhon, M.-W. Chen,
  S.~Bertolazzi, P.~Gillet, A.~{Fontcuberta i Morral}, A.~Radenovic, and
  A.~Kis.
\newblock {Large-Area Epitaxial Monolayer MoS$_{2}$}.
\newblock {\em {ACS Nano}}, 9(4):4611--4620, 2015.

\bibitem{Chen.2015b}
L.~Chen, B.~Liu, M.~Ge, Y.~Ma, A.~N. Abbas, and C.~Zhou.
\newblock {Step-Edge-Guided Nucleation and Growth of Aligned WSe$_{2}$ on
  Sapphire via a Layer-over-Layer Growth Mode}.
\newblock {\em {ACS Nano}}, 9(8):8368--8375, 2015.

\bibitem{Liu.2016}
X.~Liu, I.~Balla, H.~Bergeron, G.~P. Campbell, M.~J. Bedzyk, and M.~C. Hersam.
\newblock {Rotationally Commensurate Growth of MoS$_{2}$ on Epitaxial
  Graphene}.
\newblock {\em {ACS Nano}}, 10(1):1067--1075, 2016.

\bibitem{Chen.2015}
W.~Chen, J.~Zhao, J.~Zhang, L.~Gu, Z.~Yang, X.~Li, H.~Yu, X.~Zhu, R.~Yang,
  D.~Shi, X.~Lin, J.~Guo, X.~Bai, and G.~Zhang.
\newblock {Oxygen-Assisted Chemical Vapor Deposition Growth of Large
  Single-Crystal and High-Quality Monolayer MoS$_{2}$}.
\newblock {\em {J. Am. Chem. Soc.}}, 137(50):15632--15635, 2015.

\bibitem{Cain.2016}
J.~D. Cain, F.~Shi, J.~Wu, and V.~P. Dravid.
\newblock {Growth Mechanism of Transition Metal Dichalcogenide Monolayers: The
  Role of Self-Seeding Fullerene Nuclei}.
\newblock {\em {ACS Nano}}, 10(5):5440--5445, 2016.

\bibitem{Zhu.2017}
D.~Zhu, H.~Shu, F.~Jiang, D.~Lv, V.~Asokan, O.~Omar, J.~Yuan, Z.~Zhang, and
  C.~Jin.
\newblock {Capture the Growth Kinetics of CVD Growth of Two-Dimensional
  MoS$_{2}$}.
\newblock {\em {npj 2D Mater. Appl.}}, 1(1):8, 2017.

\bibitem{Zhou.2018}
J.~Zhou, J.~Lin, X.~Huang, Y.~Zhou, Y.~Chen, J.~Xia, H.~Wang, Y.~Xie, H.~Yu,
  J.~Lei, {Di Wu}, F.~Liu, Q.~Fu, Q.~Zeng, C.-H. Hsu, C.~Yang, L.~Lu, T.~Yu,
  Z.~Shen, H.~Lin, B.~I. Yakobson, Q.~Liu, K.~Suenaga, G.~Liu, and Z.~Liu.
\newblock {A Library of Atomically Thin Metal Chalcogenides}.
\newblock {\em {Nature}}, 556(7701):355--359, 2018.

\bibitem{Pollmann.2018}
E.~Pollmann, P.~Ernst, L.~Madau{\ss}, and M.~Schleberger.
\newblock {Ion-Mediated Growth of Ultra Thin Molybdenum Disulfide Layers on
  Highly Oriented Pyrolytic Graphite}.
\newblock {\em {Surf. Coat. Technol.}}, 349:783--786, 2018.

\bibitem{Zhang.2019}
X.~Zhang, F.~Zhang, Y.~Wang, D.~S. Schulman, T.~Zhang, A.~Bansal, N.~Alem,
  S.~Das, V.~H. Crespi, M.~Terrones, and J.~M. Redwing.
\newblock {Defect-Controlled Nucleation and Orientation of WSe$_{2}$ on hBN: A
  Route to Single-Crystal Epitaxial Monolayers}.
\newblock {\em {ACS Nano}}, 13(3):3341--3352, 2019.

\bibitem{Pollmann.2020}
E.~Pollmann, J.~M. Morbec, L.~Madau{\ss}, L.~Br{\"o}ckers, P.~Kratzer, and
  M.~Schleberger.
\newblock {Molybdenum Disulfide Nanoflakes Grown by Chemical Vapor Deposition
  on Graphite: Nucleation, Orientation, and Charge Transfer}.
\newblock {\em {J. Phys. Chem. C}}, 124(4):2689--2697, 2020.

\bibitem{Okada.2020}
M.~Okada, M.~Maruyama, S.~Okada, J.~H. Warner, Y.~Kureishi, Y.~Uchiyama,
  T.~Taniguchi, K.~Watanabe, T.~Shimizu, T.~Kubo, M.~Ishihara, H.~Shinohara,
  and R.~Kitaura.
\newblock {Microscopic Mechanism of Van der Waals Heteroepitaxy in the
  Formation of MoS$_{2}$/hBN Vertical Heterostructures}.
\newblock {\em {ACS Omega}}, 5(49):31692--31699, 2020.

\bibitem{Pollmann.2020b}
E.~Pollmann, L.~Madau{\ss}, S.~Schumacher, U.~Kumar, F.~Heuvel, C.~{vom Ende},
  S.~Yilmaz, S.~G{\"u}ng{\"o}rm{\"u}s, and M.~Schleberger.
\newblock {Apparent Differences between Single Layer Molybdenum Disulphide
  Fabricated via Chemical Vapour Deposition and Exfoliation}.
\newblock {\em {Nanotechnology}}, 31(50):505604, 2020.

\bibitem{Zhang.2019b}
F.~Zhang, Y.~Wang, C.~Erb, K.~Wang, P.~Moradifar, V.~H. Crespi, and N.~Alem.
\newblock {Full Orientation Control of Epitaxial MoS$_{2}$ on hBN Assisted by
  Substrate Defects}.
\newblock {\em {Phys. Rev. B}}, 99(15):155430, 2019.

\bibitem{Zhou.2013}
H.~Zhou, F.~Yu, Y.~Liu, X.~Zou, C.~Cong, C.~Qiu, T.~Yu, Z.~Yan, X.~Shen,
  L.~Sun, B.~I. Yakobson, and J.~M. Tour.
\newblock {Thickness-Dependent Patterning of MoS$_{2}$ Sheets with
  Well-Oriented Triangular Pits by Heating in Air}.
\newblock {\em {Nano Res.}}, 6(10):703--711, 2013.

\bibitem{Wu.2013}
J.~Wu, H.~Li, Z.~Yin, H.~Li, J.~Liu, X.~Cao, Q.~Zhang, and H.~Zhang.
\newblock {Layer Thinning and Etching of Mechanically Exfoliated MoS$_{2}$
  Nanosheets by Thermal Annealing in Air}.
\newblock {\em {Small}}, 9(19):3314--3319, 2013.

\bibitem{Yamamoto.2013}
M.~Yamamoto, T.~L. Einstein, M.~S. Fuhrer, and W.~G. Cullen.
\newblock {Anisotropic Etching of Atomically Thin MoS$_{2}$}.
\newblock {\em {J. Phys. Chem. C}}, 117(48):25643--25649, 2013.

\bibitem{Lv.2017}
D.~Lv, H.~Wang, D.~Zhu, J.~Lin, G.~Yin, F.~Lin, Z.~Zhang, and C.~Jin.
\newblock {Atomic Process of Oxidative Etching in Monolayer Molybdenum
  Disulfide}.
\newblock {\em {Sci. Bull.}}, 62(12):846--851, 2017.

\bibitem{Maguire.2019}
P.~Maguire, J.~Jadwiszczak, M.~O'Brien, D.~Keane, G.~S. Duesberg, N.~McEvoy,
  and H.~Zhang.
\newblock {Defect-Moderated Oxidative Etching of MoS$_{2}$}.
\newblock {\em {J. Appl. Phys.}}, 126(16):164301, 2019.

\bibitem{Yao.2020}
K.~Yao, J.~D. Femi-Oyetoro, S.~Yao, Y.~Jiang, L.~{El Bouanani}, D.~C. Jones,
  P.~A. Ecton, U.~Philipose, M.~{El Bouanani}, B.~Rout, A.~Neogi, and J.~M.
  Perez.
\newblock {Rapid Ambient Degradation of Monolayer MoS$_{2}$ after Heating in
  Air}.
\newblock {\em {2D Mater.}}, 7(1):015024, 2020.

\bibitem{Kaupmees.2020}
R.~Kaupmees, P.~Walke, L.~Madau{\ss}, A.~Maas, E.~Pollmann, M.~Schleberger,
  M.~Grossberg, and J.~Krustok.
\newblock {The Effect of Elevated Temperatures on Excitonic Emission and
  Degradation Processes of WS$_{2}$ Monolayers}.
\newblock {\em {Phys. Chem. Chem. Phys.}}, 22(39):22609--22616, 2020.

\bibitem{Cullen.2021}
C.~P. Cullen, O.~Hartwig, C.~Ó~Coileáin, J.~B. McManus, L.~Peters, C.~Ilhan,
  G.~S. Duesberg, and N.~McEvoy.
\newblock {Synthesis and Thermal Stability of TMD Thin Films: A Comprehensive
  XPS and Raman Study}.
\newblock {arXiv:2106.07366}, 2021.

\bibitem{Lin.2015}
J.~Lin, S.~T. Pantelides, and W.~Zhou.
\newblock {Vacancy-Induced Formation and Growth of Inversion Domains in
  Transition-Metal Dichalcogenide Monolayer}.
\newblock {\em {ACS Nano}}, 9(5):5189--5197, 2015.

\bibitem{Chen.2018}
Q.~Chen, H.~Li, S.~Zhou, W.~Xu, J.~Chen, H.~Sawada, C.~S. Allen, A.~I.
  Kirkland, J.~C. Grossman, and J.~H. Warner.
\newblock {Ultralong 1D Vacancy Channels for Rapid Atomic Migration during 2D
  Void Formation in Monolayer MoS$_{2}$}.
\newblock {\em {ACS Nano}}, 12(8):7721--7730, 2018.

\bibitem{Longo.2017}
R.~C. Longo, R.~Addou, S.~KC, J.-Y. Noh, C.~M. Smyth, D.~Barrera, C.~Zhang,
  J.~W.~P. Hsu, R.~M. Wallace, and K.~Cho.
\newblock {Intrinsic Air Stability Mechanisms of Two-Dimensional Transition
  Metal Dichalcogenide Surfaces: Basal versus Edge Oxidation}.
\newblock {\em {2D Mater.}}, 4(2):025050, 2017.

\bibitem{KC.2015}
S.~KC, R.~C. Longo, R.~M. Wallace, and K.~Cho.
\newblock {Surface Oxidation Energetics and Kinetics on MoS$_{2}$ Monolayer}.
\newblock {\em {J. Appl. Phys.}}, 117(13):135301, 2015.

\bibitem{Komsa.2013}
H.-P. Komsa, S.~Kurasch, O.~Lehtinen, U.~Kaiser, and A.~V. Krasheninnikov.
\newblock {From Point to Extended Defects in Two-Dimensional MoS$_{2}$:
  Evolution of Atomic Structure under Electron Irradiation}.
\newblock {\em {Phys. Rev. B}}, 88(3):035301, 2013.

\bibitem{Hong.2017}
J.~Hong, Y.~Pan, Z.~Hu, D.~Lv, C.~Jin, W.~Ji, J.~Yuan, and Z.~Zhang.
\newblock {Direct Imaging of Kinetic Pathways of Atomic Diffusion in Monolayer
  Molybdenum Disulfide}.
\newblock {\em {Nano Lett.}}, 17(6):3383--3390, 2017.

\bibitem{Maksov.2019}
A.~Maksov, O.~Dyck, K.~Wang, K.~Xiao, D.~B. Geohegan, B.~G. Sumpter, R.~K.
  Vasudevan, S.~Jesse, S.~V. Kalinin, and M.~Ziatdinov.
\newblock {Deep Learning Analysis of Defect and Phase Evolution during Electron
  Beam-Induced Transformations in WS$_{2}$}.
\newblock {\em {npj Comput. Mater.}}, 5(1):12, 2019.

\bibitem{Yu.2015}
Z.~G. Yu, Y.-W. Zhang, and B.~I. Yakobson.
\newblock {An Anomalous Formation Pathway for Dislocation-Sulfur Vacancy
  Complexes in Polycrystalline Monolayer MoS$_{2}$}.
\newblock {\em {Nano Lett.}}, 15(10):6855--6861, 2015.

\bibitem{Sensoy.2017}
M.~G. Sensoy, D.~Vinichenko, W.~Chen, C.~M. Friend, and E.~Kaxiras.
\newblock {Strain Effects on the Behavior of Isolated and Paired Sulfur Vacancy
  Defects in Monolayer MoS$_{2}$}.
\newblock {\em {Phys. Rev. B}}, 95(1):014106, 2017.

\bibitem{Kang.2015}
K.~Kang, S.~Xie, L.~Huang, Y.~Han, P.~Y. Huang, K.~F. Mak, C.-J. Kim,
  D.~Muller, and J.~Park.
\newblock High-mobility three-atom-thick semiconducting films with wafer-scale
  homogeneity.
\newblock {\em Nature}, 520(7549):656--660, 2015.

\end{thebibliography}
\bibliographystyle{unsrt}
%\bibliography{Bib}

\end{document}